\documentclass[%
 reprint,
nofootinbib,
 amsmath,amssymb,
prd,
]{revtex4-1}

\usepackage{graphicx}
\usepackage{dcolumn}
\usepackage{bm}
\usepackage{amsmath,amsfonts,amssymb,amsthm,bm,upgreek,dsfont}

\begin{document}

\newcommand{\tabincell}[2]{\begin{tabular}{@{}#1@{}}#2\end{tabular}}
\def\be#1\ee{\begin{equation}\begin{split}#1\end{split}\end{equation}}
\def\({\left(}
\def\){\right)}
\def\[{\left[}
\def\]{\right]}
\def\f{\frac}
\def\bea{\begin{eqnarray}}
\def\eea{\end{eqnarray}}
\def\al{\gamma}
\def\a{\alpha}
\def\b{\beta}
\def\la{\lambda}
\def\p{\partial}
\def\phis{\phi^\ast}
\def\psis{\psi^\ast}
\def\s#1{|_{#1}}
\def\ttb{T\bar{T}}
\def\diag{\mbox{diag}}
\def\cL{\mathcal{L}}
\def\hs{\hspace}
\def\nn{\nonumber\\}
\def \arcsinh {{\rm arcsinh}}
\def\la{\langle}
\def\ra{\rangle}
\def\dagger{\ast}
\def\de{\delta}

\title{Note on the nonrelativistic $T\bar{T}$ deformation}
\author{Bin Chen$^{1,2,3}$}
 \email{bchen01@pku.edu.cn}
\author{Jue Hou$^1$}
 \email{houjue@pku.edu.cn}
\author{Jia Tian$^4$}
 \email{wukongjiaozi@ucas.ac.cn}
 \affiliation{
		$^{1}$School of Physics, Peking University, No.5 Yiheyuan Rd, Beijing 100871, P.~R.~China\\
		$^{2}$Center for High Energy Physics, Peking University, No.5 Yiheyuan Rd, Beijing 100871, P.~R.~China\\
		$^{3}$Peng Huanwu Center for Fundamental Theory, Hefei, Anhui 230026, China\\
		$^4$Kavli Institute for Theoretical Sciences (KITS),\\
		University of Chinese Academy of Science, 100190 Beijing, P.~R.~China
 }

\date{\today}

\begin{abstract}
In this paper, we present our study on the $\ttb$-deformation of non-relativistic complex scalar field theory. We find the closed form of the deformed Lagrangian by using the perturbation and the method of characteristics. Furthermore we compute the exact energy spectrum of the deformed free theory  by using the Brillouin-Wigner perturbation theory in an appropriate regularization scheme.
\end{abstract}

\maketitle


\section{\textbf{Introduction}}
Solvable irrelevant deformations have attracted many interests in recent years due to their novel features that provide analytical control on the ultraviolet (UV) physics regardless of the difficulties from strong coupling or nonlocality. $\ttb$-deformation is the most studied example of such deformations\footnote{For a pedagogical review see \cite{LectureJiang}, and for recent progress see the workshop \cite{TTbar2020}.}  \cite{Smirnov:2016lqw,Cavaglia:2016oda}. It is originally defined in two-dimensional relativistic quantum field theories where $\ttb$ is a well defined composite operator that can be written as the determinant of the energy momentum tensor operator. Many interesting physical quantities such as the energy spectrum, the partition function, the entanglement entropy and the S-matrix have been computed exactly in the deformed theories, see \cite{LectureJiang} and the references within.  These exact results reveal fruitful and meanwhile unexpected structures of solvable irrelevant deformations which we have not fully understood. For example \cite{Hage1,Hage2,Hage3}, the deformed density of states of a deformed field theory shows the Hagedorn behavior which typically happens in string theory. On the other hand, the $\ttb$-deformation to holographic conformal field theory opens a new window to study AdS/CFT correspondence \cite{verlinde,Hage1,kutasov2,dubovsky1,dubovsky2,bihdtott,ttandcf,cutoffads,commentsontt,Chen:2018eqk,Hartman:2018tkw,Caputa:2019pam,Murdia:2019fax,Jeong:2019ylz,Guica:2019nzm,Chen:2019mis,Grieninger:2019zts,Lewkowycz:2019xse,Apolo:2019zai,Hirano:2020nwq, Jafari:2019qns, He:2020hhm,Kruthoff:2020hsi}.

\par
Recently similar integrable deformations have been considered in other types of models including integrable lattice models \cite{Lattice1,Lattice2} and non-relativistic integrable field theories \cite{NonRe1,NonRe2,BoseJ}. Interestingly $\ttb$ (like) deformation can be interpreted as a special type of algebra-preserving deformations studied in the context of integrable spin chains \cite{Long1}. The standard integrability technique of solving these integrable deformed models  is first to compute the deformed S-matrix in the infinite volume limit where the deformation simply modifies the S-matrix by multiplying a 1.	Castillejo-Dalitz-Dyson-like factor \cite{Smirnov:2016lqw,CDD,CDD2}, and then to substitute the deformed S-matrix into the Bethe equation to solve the theory in the finite volume.
 Using this integrability technique the deformed one-dimensional Bose gas  known as the Lieb-Liniger model was carefully studied in \cite{BoseJ} where the author showed that the deformed one-dimensional Bose gas shares many qualitative features with the $\ttb$-deformed relativistic quantum field theories. For example, the spectrum can also be given by a flow equation and the density of states exhibits the Hagedorn behavior in the thermodynamic limit.

In this work we will use another method, the quantum perturbation theory, to study the spectrum of the deformed non-relativistic field theories which are not necessarily integrable. The model we focus on is the non-relativistic  Schr\"{o}dinger model which can be viewed as the free limit of the Lieb-Liniger model. We find that for models with a degenerate Legendre transformation such as non-relativistic  Schr\"{o}dinger or Lieb-Liniger model, the Lagrangian and Hamiltonian definition of the $\ttb$-deformation may not be equivalent. In particular, within the Hamiltonian definition of $\ttb$-deformation only the flow equation of the Hamiltonian is not enough to define the deformation, one also needs to specify how the constraints change under the flow. In the first-order deformation, the Legendre transformation is degenerate so one can use the Dirac-Bergmann algorithm to quantize the system and  compute the spectrum perturbatively. We find that the perturbative expansion of the deformed spectrum is superficially divergent as expected for the models under an irrelevant deformation. After imposing the Dirichlet regularization and dropping off the transcendental terms which are not supposed to appear in the $\ttb$-deformation, the deformed spectrum match the results derived from the flow equation. We will also show that after including the second-order deformation the theory either does not admit a perturbative description or has ambiguities.

 The paper is organized as follows: in Sect. \uppercase\expandafter{\romannumeral2}, we derive the closed form of $\ttb$-deformed Lagrangian in non-relativistic complex scalar theory; in Sect. \uppercase\expandafter{\romannumeral3}, we compute the deformed energy spectrum perturbatively, and manage to find an exact form by using Brillouin-Wigner perturbation; in Sect. \uppercase\expandafter{\romannumeral4}, we end with conclusions; we put some technical details into four appendices. 

\bigskip

{\noindent \bf{Note added:}} When we were finishing this project and had obtained the main results in this paper, two interesting papers \cite{Italy,JiangNew}
appeared on arXiv in the same day. The same closed form of a $\ttb$-deformed Lagrangian of non-relativistic models have  also been derived. In \cite{Italy}, the Lagrangian is derived from  the dynamical change of coordinates, and in \cite{JiangNew}  it is derived using a  geometric method. Our method in this work is similar to the one used in \cite{Close}, and  is somewhat complementary to the ones in\footnote{We would like to mention that the closed form of the $\ttb$-deformed Lagrangian of non-relativistic models was also found by S. Frolov et.al.  from the light-cone gauge interpretation \cite{LG}.} \cite{Italy,JiangNew}.

\section{Non-relativistic $T\bar{T}$-deformed Lagrangian}
\noindent
In this section, we consider the $T\bar{T}$-deformation of a two-dimensional (2D) non-relativistic field theory whose Lagrangian satisfies\footnote{We will work in the convention that
\be
g_{\mu\nu}=\diag(1,-1),\hs{3ex}
x^{\mu}=(x^0,x^1)=(t,x).\nn
\ee
}
\be\label{ttbareq}
\frac{\p \mathcal{L}^{(\lambda)}}{\p \lambda}=\operatorname{det}\left(T_{\mu \nu}^{(\lambda)}\right),
\ee
where $T_{\mu\nu }$ is the energy momentum tensor of this theory. Our goal is to solve this flow equation \eqref{ttbareq} for a non-relativistic complex scalar theory with a potential $V(|\phi|)$.

\subsection{One real scalar case}
\noindent
To get some insights about \eqref{ttbareq} we warm up with a toy model involved with only a free real scalar, whose  Lagrangian density reads
\be
\mathcal{L}_0=\phi\p_0 \phi+(\p_1\phi)^2.
\ee
To solve \eqref{ttbareq} we expand the Lagrangian with respect to $\lambda$:
\be\label{ExpandL}
\cL=\sum_{n=0}\lambda^{n}\cL_n.
\ee
Substituting this expansion into the flow equation \eqref{ttbareq} one can read off $\mathcal{L}_n$ term by term, and the first several orders of the Lagrangian are of the forms
\be
&\mathcal{L}_0=X+Y,\\
&\mathcal{L}_1=Y(X+Y),\\
&\mathcal{L}_2=Y(X^2+ 3XY+2Y^2),\\
&\mathcal{L}_3=Y(X^3+6X^2Y+10XY^2+5Y^3),\\
&\mathcal{L}_4=Y(X^4+10X^3Y+30X^2Y^2+35XY^3+14Y^4),\\
\ee
where we have defined the convenient variables:
\be
X\equiv \phi\p_0 \phi, \hspace{3ex}Y\equiv (\p_1\phi)^2.\ee
We observe that these terms can be cast into a general form as
\be
\mathcal{L}_n
&= Y \sum_{k=0}^{n} \f{1}{k} X^{n-k} Y^{k} C_{n-1}^{k-1}C_{n+k}^{n-1}\\
&=X^n Y {}_2F_1\(-n,n+1,2,-\f{Y}{X}\),\hspace{3ex}n=1,2,3,....
\ee
Performing the summation \eqref{ExpandL} directly gives the closed form of the Lagrangian
\be
\mathcal{L}
&=X+\sum_{k=0}^{\infty} \sum_{n=k}^{n} \lambda^n Y \f{1}{k} X^{n-k} Y^{k} C_{n-1}^{k-1}C_{n+k}^{n-1}\\
&=-\f{1}{2\lambda}\(-1-\lambda X+\sqrt{(1-\lambda X)^2-4 \lambda Y}\),
\ee
which indeed solves the flow equation \eqref{ttbareq}.

Next we add in a potential $V$ which is just a function of $\phi$ but does not contain the terms of the derivative of $\phi$. As before we expand the Lagrangian with respect to $\lambda$. The first few terms in the deformed Lagrangian are
\be
\mathcal{L}_0=&-V+X+Y\\
\mathcal{L}_1=&-V^2+VX+Y(X+Y)\\
\mathcal{L}_2=&-V^3+V^2X-VY^2+Y(X^2+ 3XY+2Y^2)\\
\mathcal{L}_3=&-V^4+V^3X-VY^2(3X+4Y)\\&+Y(X^3+6X^2Y+10XY^2+5Y^3)\\
\mathcal{L}_4=&-V^5+V^4X+2V^2Y^3-VY^2(6X^2+20XY+15Y^2)\\&+Y(X^4+10X^3Y+30X^2Y^2+35XY^3+14Y^4),\\
\ee
which can be rewritten in a general form as
\be \label{Gen2}
\mathcal{L}_n
&=-V^{n+1}+V^n X+\sum_{j=0}^{n} (-V)^{j} Y^{j+1}\sum_{k=0}^{n-2j}K(n,j,k),\\
K(n,j,k)&= X^{n-2j-k} Y^{k}C_{n+k}^{n-j-1}C_{n-j-1}^{k-1+j}C_{k-1-j}^{j}\f{1}{k}.
\ee
Substituting \eqref{Gen2} into \eqref{ExpandL} and performing the summation we end up with the closed form of the deformed Lagrangian
\be\label{1rsL}
\mathcal{L}
=&-\f{V}{1-\lambda V}\\&-\f{1}{2\lambda(1-\lambda V))}\[-1-\lambda X+\sqrt{(1-\lambda X)^2-4 \lambda(1-\lambda V) Y}\].
\ee

Alternatively the deformed Lagrangian can also be obtained by using the method of characteristics with the initial condition
\be
\mathcal{L}_0=\phi \a +\b^2-V,\quad \mbox{with}\hs{2ex}\a=\p_0 \phi,\ \b=\p_1 \phi.
\ee
In terms of these new variables $\alpha$ and $\beta$ the flow equation \eqref{ttbareq} could be  rewritten as
\be
\f{\p \mathcal{L}}{\p \lambda}=-\mathcal{L}^2+\b\mathcal{L}\f{\p \mathcal{L}}{\p \b}+\a\mathcal{L}\f{\p \mathcal{L}}{\p \a},
\ee
which is equivalent to a set of equations:
\begin{equation}\label{chara1}
\left\{
\begin{aligned}
&\f{d \lambda}{d s}=1\\
&\f{d \a}{d s}=-\a \mathcal{L}\\
&\f{d \b}{d s}=-\b \mathcal{L}\\
&\f{d \mathcal{L}}{d s}=-\mathcal{L}^2\\
&\mathcal{L}(s=0)\\&=\phi \a(0)+\b(0)^2-V
\end{aligned}
\right.
\Rightarrow\hs{2ex}
\left\{
\begin{aligned}
&\lambda=s\\
&\a=\f{C_3}{-s+C_2}\\
&\b=\f{C_4}{-s+C_2}\\
&\mathcal{L}=\f{1}{s-C_2}\\
&-\f{1}{C_2}=\phi \f{C_3}{C_2}+(\f{C_4}{C_2})^2-V.
\end{aligned}
\right.
\end{equation}
Canceling the constants $C_i$'s in \eqref{chara1}, we get
\be
\mathcal{L}
&=-\f{V}{1-\lambda V}\\&-\f{1}{2\lambda(1-\lambda V))}\[-1-\lambda X+\sqrt{(1-\lambda X)^2-4 \lambda(1-\lambda V) Y}\],
\ee
which coincides with \eqref{1rsL}.

\subsection{One complex scalar case}
\noindent
Now let us turn to the complex scalar case.
 The Lagrangian density of the free complex scalar theory is
\be
\mathcal{L}_0=\f{i}{2}(\phis\p_0 \phi-\phi\p_0 \phis)-\p_1\phi\p_1\phis
\ee
Define $\vec{\a}\equiv\(\p_0 \phi,\p_0 \phis\)$ and $\vec{\b}\equiv\(\p_1 \phi,\p_1 \phis\)$.
The $\ttb$-flow equation is given by
\be\label{chara2}
\f{\p \mathcal{L}}{\p \lambda}=&-\mathcal{L}^2+\mathcal{L}\vec{\b}\cdot\f{\p \mathcal{L}}{\p \vec{\b}}+\mathcal{L}\vec{\a}\cdot\f{\p \mathcal{L}}{\p\vec{\a}}\\&-\(\f{\p \mathcal{L}}{\p \vec{\a}}\cdot\vec{\a}\)\(\f{\p \mathcal{L}}{\p \vec{\b}}\cdot\vec{\b}\)+\(\f{\p \mathcal{L}}{\p \vec{\a}}\cdot\vec{\b}\)\(\f{\p \mathcal{L}}{\p \vec{\b}}\cdot\vec{\a}\).
\ee
Without the last two terms this flow equation \eqref{chara2} can be solved by using the method of characteristics with the initial condition
\be
\mathcal{L}(s=0)=\f{i}{2}\(\phis(\vec{\a})_1-\phi(\vec{\a})_2\)-(\vec{\a})_1(\vec{\b})_2
\ee
in exactly the same way as we described in last section.
The solution is
\be\label{ttbarL0}
\mathcal{L}
&=-\f{1}{2\lambda}\[-1-\lambda X+\sqrt{(1-\lambda X)^2-4 \lambda Y}\],
\ee
where $X=\f{i}{2}(\phis\p_0 \phi-\phi\p_0 \phis)$ and $Y=-\p_1\phi\p_1\phis$.

To include the modifications of the last two terms in \eqref{chara2}, we make an ansatz on the exact Lagrangian and expand it with respect to $\lambda$ . By matching the expanded terms\footnote{The explicit forms of the Lagrangians $\cL_i$ are quite involved, and we  list them in Appendix A. } to the first few orders we can fix our ansatz completely, and in the end we check our result with the flow equation. The final result we get is
\be
\mathcal{L}\label{Lag}
&=-\f{1}{2\lambda}\[-1-\lambda X+\sqrt{(1-\lambda X)^2-4 \lambda Y+8i \lambda^2 A-4\lambda^3 B}\]
\ee
where
\be
&A=(\phis \p_1 \phi+\phi \p_1 \phis)(\p_1 \phi\p_0 \phis-\p_0 \phi\p_1 \phis),\\
&B=\phi \phis(\p_1 \phi\p_0 \phis-\p_0 \phi\p_1 \phis)^2.\\
\ee
In a similar way, we can find the deformed Lagrangian in the case that the complex scalar has a  potential $V$,
\be\label{ttbarL}
&\mathcal{L}
=-\f{V}{1-\lambda V}-\f{1}{2\lambda(1-\lambda V)}(-1-\lambda X+\\&\sqrt{(1-\lambda X)^2-4 \lambda (1-\lambda V)Y+8i \lambda^2(1-\lambda V)A-4\lambda^3(1-\lambda V)B}).
\ee
The detailed derivation is straightforward but too tedious to be presented here.

\section{Deformed energy from quantum perturbative theory}
\noindent
In this section, we will compute the spectrum of  the $T\bar{T}$-deformed non-relativistic free boson (an effective  Schr\"{o}dinger model). The model we are considering is the free complex scalar defined on a compact region $x\in [0,R]$,  whose  Lagrangian density and Hamiltonian are respectively
\be
&\mathcal{L}=\frac{i}{2}\left(\phi^{\ast} \partial_{t} \phi-\partial_{t} \phi^{\ast} \phi\right)-\partial_{x} \phi^{\ast} \partial_{x} \phi,\label{Lag0}\\
&H=\int \mathrm{d} x \,\partial_{x} \phi^{\ast}(x) \partial_{x} \phi(x).
\ee
This simple free theory can be though as the free boson limit of the Lieb-Liniger model whose $\ttb$-deformation has been well studied in \cite{BoseJ}.
The momentum operator is
\begin{equation}P=-\frac{i}{2} \int\left[\phi^{\ast}(x) \partial_{x} \phi(x)-\partial_{x} \phi^{\ast}(x) \phi(x)\right].\end{equation}
The complex scalar field satisfies the equal-time commutation relations,
\begin{equation}\begin{split}\label{commutation}
&[\phi(x, t), \phi(y, t)]=0, \quad\left[\phi^{\ast}(x, t), \phi^{\ast}(y, t)\right]=0,\\&  \left[\phi(x, t), \phi^{\ast}(y, t)\right]=\delta(x-y).
\end{split}\end{equation}
The Hilbert space is spanned by $N$-particle states,
\begin{equation}|\vec{x}\rangle=\phi^{\ast}\left(x_{1}\right) \ldots \phi^{\ast}\left(x_{N}\right)|0\rangle.\end{equation}
Here, $\vec{x}=(x_1,x_2,...,x_N)$ and $|0\rangle$ is the vacuum of Fock space, which is annihilated by $\phi(x)$. The $N$-particle eigenfunction  of this free model is simply given by the plane waves
\be&\langle \vec{x}|\vec{u}\rangle=\psi_{N}(\vec{u} \mid \vec{x})=\frac{1}{\sqrt{N !}} \sum_{\sigma \in \mathrm{S}_{N}} \exp \left[i \sum_{j=1}^{N} x_{j} u_{\sigma_{j}}\right],\\& \left|\vec{u}_{N}\right\rangle=\frac{1}{\sqrt{N !}} \int \mathrm{d}^{N} x \psi_{N}(\vec{u} \mid \vec{x})|\vec{x}\rangle,\label{Tran}\ee
with the eigenvalues
\begin{equation}E_{N}(\vec{u})=\sum_{j=1}^{N} u_{j}^{2}, \quad P_{N}(\vec{u})=\sum_{j=1}^{N} u_{j},\end{equation}
and
$$
\langle\vec{x}|\vec{y}\rangle=\sum_{\sigma \in \mathrm{S}_{N}} \delta(x_i-y_{\sigma_i}).
$$

\subsection{Lagrangian vs. Hamiltonian definition}
The $\ttb$-deformation is formally defined as a flow of the Lagrangian density of the theory:
\be
\p_\lambda \mathcal{L}=-\epsilon^{ab}\epsilon^{cd}T_{ac}^{(\lambda)}T_{bd}^{(\lambda)}\equiv-\mathcal{O}^{(\lambda)}_{T\bar{T}}
\ee
where $T_{ab}^{(\lambda)}$ is the stress tensor of the theory with the flow parameter $\lambda$. It is generally believed that at least in the classical level the deformation can be equivalently defined as a flow of the Hamiltonian density of theory:
\be \label{Hflow}
\p_\lambda \mathcal{H}=\mathcal{O}^{(\lambda)}_{T\bar{T}}.
\ee
The classical equivalence has been examined in the leading order in \cite{Kruthoff:2020hsi} and more generally in \cite{Jorjadze:2020ili}. However the equivalence is built on an implicit assumption that the Lagrangian is not singular or equivalently the Legendre transformation is invertible such that the $\ttb$-deformation and the Legendre transformation are commutative. The subtlety of a singular Lagrangian is that it leads to a constrained Hamiltonian. In this section we want to show that  in the Hamiltonian formalism the flow equation \eqref{Hflow} itself is not enough  to defined the $\ttb$-deformation.

To illustrate our point,  let us consider the simple model \eqref{Lag0} with the Lagrangian
\be
\mathcal{L}=&\mathcal{L}_0+\lambda \mathcal{L}_1,
\ee
where the leading-order deformation of the Lagrangian is
\be
\mathcal{L}_1=\frac{i}{2}\(\phi \p_t\phi (\p_x \phi^\dagger )^2-\phi^\dagger \p_t \phi^\dagger (\p_x\phi)^2\)+\(\p_x\phi^\dagger \p_x\phi\)^2.
\ee
The energy density is now:
\be
T_{00}=&\frac{\p \mathcal{L}}{\p(\p_t \phi)}\p_t \phi+\frac{\p \mathcal{L}}{\p(\p_t \phi^\dagger)}\p_t \phi^\dagger-\mathcal{L}\\
=&\p_x\phi\p_x\phi^\dagger-\lambda\(\p_x\phi^\dagger \p_x\phi\)^2.\label{T00}
\ee
This energy density is supposed to be identified with Hamiltonian density $\mathcal{H}$ after rewriting $\p_t\phi,~\p_t\phi^\dagger$ in terms of $\phi,~\phi^\dagger$ and the canonical momenta $\Pi, \Pi^\dagger$, which are given by
\be \label{Momen1}
&\Pi=\frac{\p \mathcal{L}}{\p (\p_t \phi)}=\frac{i}{2}\phi^\dagger+\frac{i\lambda}{2}\phi (\p_x \phi^\dagger)^2,\\& \Pi^\dagger=\frac{\p \mathcal{L}}{\p (\p_t \phi^\dagger)}=-\frac{i}{2}\phi-\frac{i\lambda}{2}\phi^\dagger(\p_x\phi)^2.
\ee
The problem is that the Lagrangian density $\mathcal{L}$ is singular so that the rewriting can not be performed.
Since the energy density \eqref{T00} does not depend on $\p_t\phi,~\p_t\phi^\dagger$ explicitly,  one  tends to claim
\be \label{HT00}
\mathcal{H}=T_{00}=\p_x\phi\p_x\phi^\dagger-\lambda\(\p_x\phi^\dagger \p_x\phi\)^2.
\ee
However clearly \eqref{HT00} is not consistent with the definition \eqref{Hflow} even in the leading order
\be \label{HTt}
&\mathcal{H}=\mathcal{H}_0+\lambda \mathcal{O}^{(\lambda)}_{T\bar{T}}\\&=\p_x\phi \p_x\phi^\dagger-\lambda\(\p_x\phi^\dagger \p_x\phi \)^2-\frac{i\lambda}{2} \(\phi \p_t\phi (\p_x \phi^\dagger )^2-\phi^\dagger \p_t \phi^\dagger (\p_x\phi)^2\).
\ee
Strictly speaking, this is not the expression of Hamiltonian density because it contains the time-derivative terms $\p_t\phi,~\p_t\phi^\dagger$ which come from $T_{ab}$. Using the equations of motion
\be
i\p_t \phi+\p_x^2\phi=i\p_t \phi^\dagger-\p_x^2\phi^\dagger=0+\mathcal{O}(\lambda)
\ee
one can rewrite \eqref{HTt} as
\be \label{HT001}
&\mathcal{H}=\p_x\phi \p_x\phi^\dagger-\lambda\(\p_x\phi^\dagger \p_x\phi \)^2\\&+\frac{\lambda}{2}\((\p_x\phi^\dagger)^2\p_x^2\phi \phi+(\p_x\phi)^2\p_x^2\phi^\dagger\phi^\dagger\).
\ee
It is obvious that \eqref{HT00} and \eqref{HT001} are in mismatch. This mismatch  is problematic when one tries to compute the spectrum of the theory perturbatively.  This problem does not appear in a relativistic theory, for example in a theory of the relativistic free boson. In the relativistic case,  the canonical momentum depends on the flow parameter as well such that the deformed Hamiltonian densities are the same at the classical level in two different ways,  as we show in the Appendix \ref{Diff}.  A more severe problem about the definition \eqref{Hflow} is that it is not enough to define the deformation. For example if want to obtain the second-order correction of \eqref{HT001} we need to first introduce the Lagrangian density
\be \label{Lan}
\mathcal{L}=\p_t\phi \Pi+\p_t^\dagger\phi \Pi^\dagger-\mathcal{H},
\ee
with the primary constraints
\be \label{Constr}
\chi_1=\Pi-\frac{i}{2}\phi^\dagger=0,\quad \chi_2=\Pi^\dagger+\frac{i}{2}\phi=0,
\ee
then derive the first-order canonical stress tensor through
\be
T^a_b=\frac{\p \mathcal{L}}{\p(\p_a \phi)}\p_a \phi+\frac{\p \mathcal{L}}{\p(\p_a \phi^\dagger)}\p_a \phi^\dagger-\delta^a_b \mathcal{L},
\ee
and in the end substitute it into \eqref{Hflow}. Therefore besides the flow equation \eqref{Hflow} one also needs to know how the constraints $\chi_1^{(\lambda)}$ and $\chi_2^{(\lambda)}$ change under the flow.\par
Assuming that we use \eqref{Hflow} together with $\chi_1^{(\lambda)}$ and $\chi_2^{(\lambda)}$ to define the $\ttb$-deformation of the non-relativistic free boson gas,  we can use the Hellmann-Feynman theorem and factorization formula  to derive the flow equation of the spectrum
\be\label{floweqn}
\frac{d \langle n| H |n\rangle}{d\lambda}=\langle n|T_{00}|n\rangle \langle n|T_{11}|n\rangle -\langle n|T_{01}|n\rangle \langle n|T_{10}|n\rangle .
\ee
But because the mismatch of \eqref{HT00} and \eqref{HT001} we can not identify $\la n|T_{00}|n\ra$ with $E_n/R$ and $\la n| H |n\ra=E_n$ at the same time if we use  \eqref{HT001} as the Hamiltonian density, where $R$ is the size of the system.

\par
It is not hard to check that the two constraints \eqref{Constr} are the second-class constraints so one may wonder how about to solve them to reduce the dimensions of the phase space first and then perform the $\ttb$-deformation. The reduced Hamiltonian density is given in \cite{Gergely:2003zy}
\be \label{reduce}
\mathcal{H}_r=-\frac{i}{2}\p_x \tilde{\phi} \p_x \tilde{\Pi},\quad \tilde{\phi}=\frac{\phi}{2}+i\Pi^\dagger,\quad \tilde{\Pi}=\Pi+i\frac{\phi^\dagger}{2}.
\ee
However this does not solve the problem: we still can not transform the Hamiltonian density into a Lagrangian density which only depends on $\tilde{\phi}$ and $\p_t \tilde{\phi}$ so that the $\ttb$-deformation can not be defined. A caveat of our analysis is that there may exist other ways to describe the reduced Hamiltonian system such that the velocity $\p_t \phi$ could be solved in terms of the momentum.\par

The singular Lagrangian density can be cured by including the higher-order deformations. In particular the exact Lagrangian is not singular so in principle we can use the Legendre transformation to obtain a Hamiltonian density without any constraints. But it turns out the resulting Hamiltonian has a singularity at $\lambda=0$, indicating that it can not be expanded in a Taylor series of the flow parameter $\lambda$. Requiring  the vanishing of these singularities introduces the constraints. Since there are different ways to introduce the constraints,  there will be ambiguities in the constrained Hamiltonian. In short, the mismatch \eqref{HT00} and \eqref{HT001}  is not just a problem in the first-order deformation, instead it is a problem of the definition of the $\ttb$-deformation in a system with singular Lagrangian density or equivalently in a constrained Hamiltonian system.

We conclude this section by stressing that  the above argument is purely classical, without considering  the problem of Feynman path integral and the ambiguities in the operator ordering.

\subsection{Dirac-Bergmann Algorithm}
The standard technique to solve the constrained Hamiltonian system is through the Dirac-Bergman Algorithm (DBA).
In this section, we will focus on the first-order deformation and study how the deformation affects the commutation relations using DBA.
First let us demonstrate the algorithm in the free theory with the action \eqref{Lag0}.
The conjugate momentum of $\phi$ and $\phi^\dagger$ are
\be
\Pi=\frac{i}{2}\phi^\dagger,\quad \Pi^\dagger=-\frac{i}{2}\phi,
\ee
satisfying the standard equal-time commutation relation
\be
[\phi(x),\Pi(y)]=i\delta(x-y),\quad [\phi^\dagger(x),\Pi^\dagger(y)]=i\delta(x-y).
\ee
The Hamiltonian system has two constraints \eqref {Constr}
satisfying
\be
[\chi_1,\chi_2]\equiv C_{1_x,2_y}=\delta(x,y).
\ee
These two constraints are the second class, so we can introduce the Dirac bracket which is defined by
\be
&[\phi(x),\phi^\dagger(y)]_D\\
&=[\phi(x),\phi^\dagger(y)]-\int dx' dy'[\phi(x),\chi_a(x')]C_{a_x',b_y'}^{-1}[\chi_b(y'),\phi^\dagger(y)]\\
&=0-\int dx' dy'\(i\delta(x-x')(-\delta(x'-y'))(-i\delta(y'-y))\)\\
&=\delta(x-y)
\ee
as expected, recalling that   the inverse of the matrix $C^{-1}_{a,b}$ is defined as
\be \label{InC}
\int dx' C_{a_x,b_x'}C^{-1}_{b_x',c_y}=\delta_{ac}\delta(x-y).
\ee
Now we consider the first-order deformed theory with the action
\be
&\mathcal{L}=\mathcal{L}_0+\lambda \mathcal{L}_1,\\
&\mathcal{L}_1=(\p_x\phi^\dagger \phi_x \phi)^2+\frac{1}{2}\(i\phi \p_t\phi (\p_x\phi^\dagger)^2-i\phi^\dagger\p_t\phi^\dagger(\p_x\phi)^2\).\label{FirstLag}
\ee
Correspondingly the two constraints become
\be
&\chi_1=\Pi-\frac{i}{2}\phi^\dagger-\frac{i\lambda}{2}\phi (\p_x\phi^\dagger)^2,\\& \chi_2=\Pi^\dagger+\frac{i}{2}\phi+\frac{i\lambda}{2}(\phi^\dagger (\p_x\phi)^2).
\ee
From them we can compute
\be
[\chi_1(x),\chi_2(y)]&\equiv C_{1_x,2_y}=-C_{2_y,1_x}\\
&=\delta(x-y)+\lambda G(x,y).
\ee
where
\be
G(x,y)=\p_y\delta(x-y)\phi^\dagger(y)\p_y\phi(y)+\p_x\delta(x-y)\p_x\phi^\dagger(x)\phi(x).
\ee
The inverse of the matrix $C_{1_x,2_y}$ up to the first order of $\lambda$ is simply
\be
C_{1_x,2_y}^{-1}=-\delta(x-y)+\lambda G(y,x).
\ee
Then at the first order of $\lambda$ the non-vanishing commutation relations between $\phi$ and $\phi^\dagger$ is
\be
[\phi(x),\phi^\dagger(y)]=0-C_{1_x,2_y}^{-1}=\delta(x-y)-\lambda G(y,x).
\ee
This implies that $\phi$ and $\phi^\dagger$ are not canonical variables so that $\phi^\dagger$ is not the  creation operator to generate the Fock space.  This fact is a reflection of the fact that $T\bar{T}$-deformation is equivalent to a dynamical coordinate transformation.

At the leading order we can introduce the canonical variables
\be\label{NewF}
\varphi=\phi+\frac{\lambda }{2} \phi^\dagger (\p_x \phi)^2,\quad \varphi^\dagger=\phi^\dagger+\frac{\lambda }{2} (\p_x \phi^\dagger)^2 \phi ,\ee
and conversely
\be
 \phi=\varphi-\frac{\lambda}{2} \varphi^\dagger(\p_x\varphi)^2,\quad \phi^\dagger=\varphi^\dagger-\frac{\lambda}{2}\varphi(\p_x\varphi^\dagger)^2,
\ee
such that
\be
\la x_1 x_2|x_3 x_4\ra=\la \varphi_1 \varphi_2 \varphi_3^\dagger \varphi_4^\dagger \ra=\delta(x_{13})\delta(x_{24})+\delta(x_{14})\delta(x_{23}).
\ee
The relations \eqref{NewF} are not canonical transformation and they are \textit{not} the state flow which was introduced in \cite{Kruthoff:2020hsi}.  In terms of the new canonical variables the Hamiltonian density \eqref{HT00} up to the first order reads
\be \label{H1b}
T_{00}=\mathcal{H}\equiv H_0+ H_1+\mathcal{O}(\lambda^2)
\ee
with
\be
H_0=&\p_x\varphi\p_x\varphi^\dagger\\
H_1=&-2\lambda (\p_x\varphi \p_x\varphi^\dagger)^2\\&-\lambda\(\varphi\p_x\varphi \p_x\varphi^\dagger \p_x^2\varphi^\dagger+\varphi^\dagger \p_x\varphi^\dagger\p_x\varphi\p_x^2\varphi \).
\ee
As a consistency check we also work out the first-order deformation of the number operator and momentum operator
\be
N&=-\frac{i}{2} \( \frac{\p \mathcal{L}}{\p \p_t \phi}\phi- \frac{\p \mathcal{L}}{\p \p_t \phi^\dagger}\phi^\dagger\)\\&=\phi^\dagger \phi+\frac{\lambda}{2}\({\phi^\dagger}^2(\p_x\phi)^2+(\p_x\phi^\dagger)^2\phi^2 \)\\
&=\varphi^\dagger \varphi+\mathcal{O}(\lambda^2),
\ee
\be
P&=\frac{\p \mathcal{L}}{\p \p_t \phi}\p_x\phi+\frac{\p \mathcal{L}}{\p \p_t \phi^\dagger}\p_x\phi^\dagger\\&=\frac{i}{2}(\p_x\phi \phi^\dagger-\p_x \phi^\dagger\phi)(1-\lambda \, \p_x\phi\p_x\phi^\dagger),\quad\\
&=\frac{i}{2}(\p_x\varphi \varphi^\dagger-\p_x \varphi^\dagger\varphi)+\text{total derivatives}+\mathcal{O}(\lambda^2)
\ee
which are indeed undeformed as expected. We expect that the spectrum of \eqref{H1b} can be derived by performing the second quantization and applying the standard quantum perturbation theory.

\subsection{Two-particle sector without momentum}
The explicit form of the Hamiltonian \eqref{H1b} implies that there is only two-to-two scattering,  so without losing any generality we can focus on the two particle sector. When the total momentum is zero, the $\ttb$-deformed spectrum satisfies a very simple equation \cite{BoseJ}
\be
E_N(R,\lambda)=E_N(R+\lambda E_N,0).
\ee
To illustrate the procedure of our calculation, we start from the two-particle sector with zero momentum, $u_1=-u_2=u>0$.  In this case, a general two-particle eigenfunction is
\be\label{eiginfunction0}
\psi=\f{\sqrt{2}}{R} \cos\(u(x_1-x_2)\),
\ee
with
\be\label{Betheroot}
u_I=\f{2\pi}{R}I,\quad I\in\mathbf{Z}^+.
\ee

First we compute the matrix elements of  the normal ordered Hamiltonian \eqref{H1b} of the field theory in the position space
\be
\langle \vec{y}|H_0+ H_1 |\vec{x}\rangle.
\ee
For example the undeformed Hamiltonian density $\mathcal{H}_0$ is evaluated to be
\be
\langle \vec{y}|\mathcal{H}_0|\vec{x}\rangle&=\langle0|\varphi(y_1)\varphi(y_2) \p_x\varphi^\dagger \p_x\varphi \varphi^\dagger(x_1)\varphi^\dagger(x_2)|0\rangle\\
&=\p_x\delta(x-x_1)\p_x\delta(x-y_2)\delta(y_1-x_2)\\&+\p_x\delta(x-x_2)\p_x\delta(x-y_2)\delta(y_1-x_1)\\
&+\p_x\delta(x-x_1)\p_x\delta(x-y_1)\delta(y_2-x_2)\\&+\p_x\delta(x-x_2)\p_x\delta(x-y_1)\delta(y_2-x_1).\\
\ee
Then we use the wavefunction \eqref{Tran} to obtain the Hamiltonian
\be\label{cal1}
H_{0II}
&\equiv \langle u_I|H_0 |u_I\rangle=\frac{1}{2}\int d\vec{y\,}d\vec{x}\,dx \,\psi_I^\ast (\vec{y})\psi_I(\vec{x})\langle \vec{y}|H_0|\vec{x}\rangle\\
&=\int d\vec{x}\,\psis_I(\vec{x})(-\p^2_{x_1}-\p^2_{x_2})\psi_I(\vec{x})\\
&=2 u_I^2\equiv \al I^2,
\ee
where $\al=\frac{8\pi^2}{R^2}$.
Similarly we find
\be \label{H1pmn}
H_{1IJ}=-\frac{\lambda}{2} (\frac{\sqrt{2}}{R})^2\left(8 u_I^2 u_J^2 R)\right)=-2\lambda \frac{\al^2 I^2 J^2}{R}.
\ee
 The result $H_{1II}=-2\lambda/ R H_{0II}^2$  coincides with the subleading term in the $\lambda$ expansion of the \textit{exact} spectrum:
\be \label{Exact}
&E_N(R,\lambda)=E_N^{(0)}-\frac{2\lambda}{R}{E_N^{(0)}}^2+\frac{7\lambda^2}{R^2}{E_N^{(0)}}^3-\frac{30\lambda^3}{R^3}{E_N^{(0)}}^4\\&+\frac{143 \lambda^4}{R^4}{E_N^{(0)}}^5-\frac{728\lambda^5}{R^5}{E_N^{(0)}}^6\dots
\ee
which is solved from the flow equation in \cite{BoseJ}.

\subsection*{Rayleigh-Schr\"{o}dinger perturbation}
The higher-order perturbative results can be obtained by using the Rayleigh-Schr\"{o}dinger perturbation:
\be \label{RSE}
E^{(2)}_I=&I^4(-2\lambda)^2 \f{\al^3}{R^2} g_I(1),\\
E^{(3)}_I=&I^4(-2\lambda)^3 \f{\al^4}{R^3}\(g_I(1)^2-g_I(2)\)\\
E^{(4)}_I=&I^4(-2\lambda)^4 \f{\al^5}{R^4}\(g_I(1)^3-3g_I(1)g_I(2)+g_I(3)\)\\
E^{(5)}_I=&I^4(-2\lambda)^5 \f{\al^6}{R^5}\(g_I(1)^4-6g_I(1)^2g_I(2)\right.\\&\left.+2g_I(2)^2+4g_I(1)g_I(3)-g_I(4)\)\\
&....\nonumber
\ee
where for convenience we have introduced the function
\be
g_I(a)\equiv \sum_{J\neq I}^{\infty}\f{I^{4(a-1)}J^4}{(I^2-J^2)^a}.
\ee
The functions $g_I(1)$ and $g_I(2)$ are divergent, and proper regularization has to be adopted. The first few terms of $g_I(a)$ are given by
\be
&g_I(1)=\f{7I^2}{4},\quad g_I(2)=-\f{11I^4}{16}+\f{\pi^2I^6}{12},\\& g_I(3)=-\f{I^6}{32}-\f{5\pi^2I^8}{48},\\
&g_I(4)=-\f{3I^8}{256}+\f{\pi^2I^{10}}{96}+\f{\pi^4I^{12}}{720},....
\ee
where we have used the Dirichlet regularization to regularize $g_I(1)$ and $g_I(2)$. For example, to compute $g(1)$ we first separate it into the divergent piece and convergent piece
\be
g_I(1)=\sum_{J\neq I}^{\infty}\([-I^2-J^2]+\f{I^4}{I^2-J^2}\)
\ee
then regularize the divergent summation with the Dirichlet regularization.
 Substituting the regularized functions into \eqref{RSE}, we obtain the higher-order quantum corrections to the spectrum:
\be\label{TwoSpectrum}
&E_I^{(2)}=I^6\lambda^2\frac{\gamma^3}{R^2}7,\quad E_I^{(3)}=I^{8}(-\lambda)^3 \f{\al^4}{R^3}\(30-\frac{2\pi^2I^2}{3}\),\\
&E^{(4)}_I=I^{10}(-\lambda)^4 \f{\al^5}{R^4}\({143}-\frac{26\pi^2I^2}{3}\),\\
&E^{(5)}_I=I^{12}(-\lambda)^5 \f{\al^6}{R^5}\(728-\frac{2\pi^4I^4-400\pi^2I^2}{5}\),....
\ee
Surprisingly, these expressions are in exact match with \eqref{Exact} if all the transcendental terms are discarded. It was shown in \cite{Zamolodchikov:1991vx,Caselle:2013dra} that the appearance of the transcendental terms is the signal of the contributions from other irrelevant operators instead of $\ttb$, so they should be discarded. If we follow this rule, we  can use the Brillouin-Wigner perturbation to obtain the exact spectrum which coincides with the one solved from the flow equation, as we show below.

\subsection*{Exact form: Brillouin-Wigner perturbation}
\noindent
A technical introduction to the Brillouin-Wigner perturbation can be found in Appendix C.  The exact energy satisfies
\be
E_k=E_k^{(0)}+\sum_{i=1}^{\infty}\f{\prod_{j=0}^{i-1}H_{1m_j m_{j+1}}}{\prod_{j=1}^{i-1}(E_{k}-E^{(0)}_{m_j})},\quad m_0=m_i=k.
\ee
In our case, $E_n^{(0)}=\al n^2,\ \ H_{1n m}=-2\lambda \f{\al^2}{R}n^2 m^2$ so that
\be
E_k=\al k^2+\al k^4 \sum_{i=1}^{\infty} \(-\f{2\al \lambda}{R}\)^i g(k,b_k)^{i-1}
\ee
where $b_k=E_k / \al$ and
\be
g(k,b_k)&=\sum_{m\neq k,m=1}^{\infty}\f{m^4}{b_k-m^2}=\f{k^4}{k^2-b_k}+\f{1}{2}b_k^{3/2}\pi \cot(\pi\sqrt{b_k}).
\ee
The equation of the energy can be recast into the following form
\be
&b_k=k^2-\f{\beta k^4}{1+\beta \(\f{k^4}{k^2-b_k}+\f{1}{2}b_k^{3/2}\pi \cot(\pi\sqrt{b_k})\)}\\
\ee
where $\beta=\f{\al \lambda}{R}$. As in the last subsection, if we discard the $\pi$-dependent terms this result will match the exact result found in \cite{BoseJ}.
Notice that we must do this carefully. Naively, it seems necessary to throw away the term $\pi \cot(\pi\sqrt{b_k})$ completely. However, because of $b_k=k^2+O(\lambda)$, we can expand the function and find
\be
&\pi \cot(\pi\sqrt{b_k})=\pi \cot(\pi(\sqrt{b_k}-k))\\&=\f{1}{\sqrt{b_k}-k}+(\mbox{$\pi$-dependent terms}).
\ee
 Then the equation of the energy becomes
\be
\f{b_k}{k^2}=1-\f{\beta k^2}{1+\beta k^2 \(\f{1}{1-b_k/k^2}+\f{1}{2}\f{(b_k/k^2)^{3/2}}{\sqrt{b_k/k^2}-1}\)}.\\
\ee
This is an algebraic equation, and is equivalent to
\be\label{burger}
\tilde{\beta}_k\tilde{b}_k^{3/2}+\sqrt{\tilde{b}_k}-1=0
\ee
where
\be
\tilde{\beta}_k=\beta k^2/2=\f{\al \lambda k^2}{2R},\hs{3ex}\tilde{b}_k=\f{E_k}{\al k^2}.
\ee
Multiplying \eqref{burger} by $\tilde{\beta}_k\tilde{b}_k^{3/2}+\sqrt{\tilde{b}_k}+1$, we  get
\be
\tilde{\beta}_k^2\tilde{b}_k^{3}+2\tilde{\beta}_k\tilde{b}_k^{2}+\tilde{b}_k-1=0,
\ee
which is the Burgers' equation  in \cite{BoseJ}. A solution of \eqref{burger} is simply
\be
\tilde{b}_k=\f{2}{3 \tilde{\beta}_k }\[\cosh \left(\frac{2}{3}\arcsinh\left(\frac{3 \sqrt{3 \tilde{\beta}_k }}{2}\right)\right)-1\].
\ee
Expanding the above relation, we can reproduce the perturbative results in the last subsection.

\subsection{Two-particle sector with momentum}
In this subsection, we consider the two-particle sector with momentum. The general two-particle eigenfunction is
\be \label{wave2}
\psi=\frac{1}{\sqrt{2}R}\(e^{i u_1 x_1+i u_2 x_2}+e^{i u_1 x_2+i u_2 x_1}\),
\ee
with zero-order eigenvalue
\be
E^{(0)}_{u_1,u_2}=u_1^2+u_2^2.
\ee
When $u_1\neq \pm u_2$, the eigenvalues have a four-fold degeneracy $E^{(0)}_{u_1,u_2}=E^{(0)}_{-u_1,u_2}=E^{(0)}_{u_1,-u_2}=E^{(0)}_{-u_1,-u_2}$. When $u_1=u_2$, the degeneracy is two-fold: $E^{(0)}_{u_1,u_2}=E^{(0)}_{-u_1,-u_2}$. The degenerate spectrum is an obstacle to apply the flow equation \eqref{floweqn} to derive the deformed spectrum because the factorization formula fails. However in this case, because the total momentum is conserved so one can focus on a sector with some fixed total momentum $u=u_1+u_2$. Within this sector the spectrum will not be degenerate anymore and the flow equation still works. The flow equation can be solved by a formal series expansion in $\lambda$. The  first few terms of the deformed spectrum are given by\footnote{There is a typo in equation $(122)$ of \cite{BoseJ}: $E_N^{(1)}=\frac{1}{R}(-2M_2^2+2M_1M_3)$.}
\be
&\mathcal{E}_{u_1,u_2}^{(1)}=\frac{\lambda}{R}2u_1u_2(u_1-u_2)^2,\\
& \mathcal{E}_{u_1,u_2}^{(2)}=-\frac{\lambda^2}{R^2}2u_1u_2(u_1-u_2)^2(u_1^2-5u_1u_2+u_2^2),\\
&\mathcal{E}_{u_1,u_2}^{(3)}=\frac{\lambda^3}{R^3}2u_1u_2(u_1-u_2)^2(u_1^2-10u_1u_2+u_2^2)\\&~~~~~~~~~~~~(u_1^2-3u_1u_2+u_2^2),\\
&\dots
\ee

Below we will try to compute the deformed spectrum within quantum perturbation theory.
In the basis of the wavefunction \eqref{wave2}, we find
\bea
\lefteqn{\la u_1,u_2|H_1|u_3,u_4\ra}\notag\\
&=&-\frac{\lambda}{R}\delta(u_1+u_2-u_3-u_4)\notag\\
&&\times\(8u_1u_2u_3u_4-(u_1+u_2)^2(u_1u_2+u_3u_4)\),
\eea
where the Dirac delta function is due to the momentum conservation.
In particular we have
\be \label{TwoMoH1mn}
E_{u_1,u_2}^{(1)}= \la u_1,u_2|H_1|u_1,u_2\ra=\frac{\lambda}{R}2u_1u_2(u_1-u_2)^2=\mathcal{E}_{u_1,u_2}^{(1)}.
\ee
 The second-order correction is given by a divergent summation\footnote{Without losing generality we have assumed $u_1\geq u_2$.}
\begin{widetext}
\be \label{TwoMoH2}
E_{u_1,u_2}^{(2)}&=\sum_{u_3,u_4} \frac{|\la u_1 u_2|H_1|u_3,u_4\ra|^2|}{u_1^2+u_2^2-u_3^2-u_4^2}\\
&=\frac{\lambda^2}{R^2}\sum_{u_3 \neq u_1,u_3=-\infty}^{\infty}\frac{-8u_1u_2(u_1+u_2-u_3)u_3+(u_1+u_2)^2(u_1u_2+(u_1+u_2-u_3)u_3)^2}{2(u_1-u_3)(u_3-u_2)}.
\ee
\end{widetext}
As before to regularize it we first separate out the divergent piece by a $1/u_3$ Taylor expansion
\be
\eqref{TwoMoH2}=&\sum_{u_3=-\infty}^{\infty} [D_0(u_1,u_2)+D_1(u_1,u_2)u_3+D_2(u_1,u_2)u_3^2] \\
&-(D_0(u_1,u_2)+D_1(u_1,u_2)u_1+D_2(u_1,u_2)u_1^2)\\&+\sum_{u_3 \neq u_1,u_3=-\infty}^{\infty} C(u_1,u_2,u_3),\ee
with
\be
&C(u_1,u_2,u_3)=\frac{2u_1^2(u_1-u_2)^4 u_2^2}{(u_1-u_3)(u_3-u_2)}.
\ee
Then using the Dirichlet regularization
\be
\sum_{u_3=-\infty}^{\infty} 1=\sum_{u_3=-\infty}^{\infty} u_3=\sum_{u_3=-\infty}^{\infty} u_3^2=0,
\ee
we find the final result of $E_{u_1,u_2}^{(2)}$
\be
\mathcal{E}_{u_1,u_2}^{(2)}=&\frac{\lambda^2}{R^2}\left(0-\(2u_1(u_1-u_2)^2u_2(u_1^2-6u_1u_2+u_2^2)\)\right.\\&\left.+2u_1^2(u_1-u_2)^2u_2^2\right).
\ee
Similarly, we can read the third-order correction
\be
E_{u_1,u_2}^{(3)}=\mathcal{E}_{u_1,u_2}^{(3)}-\pi^2\frac{2\lambda^3}{3R^3}u_1^3u_2^3(u_1-u_2)^4,
\ee
which reduces to \eqref{TwoSpectrum} by taking $u_2=-u_1$. The calculation of higher-order corrections is much involved, but the feature is very similar so we will not present the analysis here.\par

\subsection{Three-particle sector}
To analyze the spectrum of the three-particle sector one has to include the second order deformation. However as we discuss before, after including the second order deformation the theory either does not admit a perturbative description or has ambiguities. The detail of this analysis can be found in Appendix \ref{second}. In this subsection we only compute the spectrum up to the first-order deformation.
The eigenfunction of the free three-particle theory with $I=(I_1, I_2, I_3)$ is given by
\be
\psi_I(x_1,x_2,x_3)=\f{1}{\sqrt{3!R^3}}\sum_{\sigma\in S}\exp^{i\sum_{j=1}^3 x_j u_{I\sigma_j}}
\ee
As before one can find the undeformed Hamiltonian in the position space
\be
&\langle x_1,x_2,x_3|H_0|y_1,y_2,y_3\rangle\\&=\sum_{i,j\in S}\int \,d x \p_x \de(x-x_{i_1})\p_x \de(x-y_{j_1})\de(x_{i_2}-y_{j_2})\de(x_{i_3}-y_{j_3})
\ee
and in the momentum space
\be
\langle u_I|H_0|u_J\rangle=\(\f{2\pi}{R}\)^2\sum I_{i_1} J_{j_1}\de^{I_{i_1}}_{J_{j_1}}\de^{I_{i_2}}_{J_{j_2}}\de^{I_{i_3}}_{J_{j_3}},
\ee
where $S$ denotes  the permutation of three particles $(1,2,3)$, and $u_{I_1}=\f{2\pi}{R}I_1$. Therefore the zeroth-order energy is given by\footnote{Here we assume that $I_1> I_2 >I_3$. }
\be
E_I^{(0)}=\langle u_I|H_0|u_I\rangle=\(\f{2\pi}{R}\)^2(I_1^2+I_2^2+I_3^2).
\ee
Similarly, we can get the first-order result,
\be
&\langle x_1,x_2,x_3|H_1|y_1,y_2,y_3\rangle=\\&\sum_{i,j\in S}\int \,d x \de(x-x_{i_1}) \de(x-x_{i_2}) \de(x-y_{j_1}) \de(x-y_{j_2})\de(x_{i_3}-y_{j_3})\\
&~~~\(-2\p_{x_{i_1}}\p_{x_{i_2}}\p_{y_{j_1}}\p_{y_{j_2}}-\p_{x_{i_1}}\p_{x_{i_2}}^2\p_{y_{j_1}}-\p_{x_{i_1}}\p_{y_{j_1}}\p_{y_{j_2}}^2\)
\ee
and
\be
&\langle u_I|H_1|u_J\rangle\\
&=\f{1}{R}\(\f{2\pi}{R}\)^4\sum_{i,j\in S}\de^{I_{i_3}}_{J_{i_3}}\(-2 I_{i_1}I_{i_2}J_{j_1}J_{j_2}+I_{i_1}I_{i_2}^2 J_{j_1}+I_{i_1}J_{j_1}J_{j_2}^2\)
\ee
The first order correction of the energy is given by
\be
E^{(1)}_I=&\lambda \langle u_I|H_1|u_I\rangle \\=&\lambda \f{1}{R}\(\f{2\pi}{R}\)^4\left(2I_{1}I_{2}(I_{1}-I_{2})^2\right.\\&\left.+2I_{2}I_{3}(I_{2}-I_{3})^2+2I_{3}I_{1}(I_{3}-I_{1})^2\right)
\ee
It is the same as the one in \cite{BoseJ}.\par
The second-order correction of the energy should come from $\langle u_I|H_1|u_J\rangle$ and $\langle u_I|H_2|u_I\rangle$
\be
E_{I}^{(2)}=\lambda^2 \sum_{J\neq I}\f{|\langle u_I|H_1|u_J\rangle|^2}{E_{I}^{(0)}-E_{J}^{(0)}}+\lambda^2 \langle u_I|H_2|u_I\rangle.
\ee
The first part of $E_{I}^{(2)}$ is
\begin{widetext}
\be
&\sum_{J\neq I}\f{|\langle u_I|H_1|u_J\rangle|^2}{E_{I}^{(0)}-E_{J}^{(0)}}\\&
=\f{1}{R^2}\(\f{2\pi}{R}\)^6\sum_{J\neq I}\f{1}{I_1^2+I_2^2+I_3^2-J_1^2-J_2^2-J_3^2}\sum_{i_3,j_3}\de^{I_{i_3}}_{J_{i_3}}\(-8 I_{i_1}I_{i_2}J_{j_1}J_{j_2}+(I_{i_1}I_{i_2}+J_{j_1}J_{j_2})(I_{i_1}+I_{i_2}) (J_{j_1}+J_{j_2})\)^2\nonumber
\ee
\end{widetext}
 Notice that there is no term like $\de_{J_1}^{I_2}\de_{J_2}^{I_3}$ due to $I_1+I_2+I_3=J_1+J_2+J_3$ and $J \neq I$. After some manipulations, we read
\begin{widetext}
\bea
\lefteqn{\sum_{J\neq I}\f{|\langle u_I|H_1|u_J\rangle|^2}{E_{I}^{(0)}-E_{J}^{(0)}}}\\
&=&\f{1}{R^2}\(\f{2\pi}{R}\)^6\left(-2I_1I_2(I_1-I_2)(I_1^2-5 I_1 I_2+I_2^2)-2I_1I_3(I_1-I_3)(I_1^2-5 I_1 I_3+I_3^2)-2I_3I_2(I_3-I_2)(I_3^2-5 I_3 I_2+I_2^2)\right)\nonumber
\eea
\end{widetext}
Matching $E_{I}^{(2)}$ with the result in \cite{BoseJ}, it implies that
\be
\langle u_I|H_2|u_I\rangle=&2 I_1 I_2 I_3 (5 I_1^3-6 I_1^2 I_2-6 I_1 I_2^2+5 I_2^3-6 I_1^2 I_3\\&+21 I_1 I_2 I_3-6 I_2^2 I_3-6 I_1 I_3^2-6 I_2 I_3^2+5 I_3^3).
\ee

\section{Conclusion}
\noindent
In this work we have derived a closed form \eqref{ttbarL} of the Lagrangian of the $\ttb$-deformed non-relativistic model by using the perturbative computations and the method of characteristics. In addition, we computed the deformed spectrum by using two different perturbation theories. In particular, by using the Brillouin-Wigner perturbation, we managed to find the exact form of the deformed energy spectrum, which can match the one obtained in \cite{BoseJ} after some appropriate regularization.

One subtle issue in our study is on the regularization. The perturbative expansion of the deformed energy is a summation of infinite series, which is superficially divergent. We applied the zeta-function regularization and found that the results could match the ones in \cite{BoseJ} if we discard the $\pi$-dependent terms. The similar $\pi$-dependent terms, which were referred to as the terms with nonzero transcendentality, appeared also in the study on other models \cite{Zamolodchikov:1991vx, Caselle:2013dra}. They  could come from the perturbations of higher order of $\ttb$, say $T^2\bar{T}^2$.  It would be certainly interesting to understand why they appear in our perturbative analysis.

Another interesting finding is that the $\ttb$-deformation on the Lagrangian and the one on the Hamiltonian is not equivalent if the theory has a degenerate Legendre transformation. The deformation may remove the degeneracy but the resulting deformed theory would not admit a perturbative description. More importantly, we want to stress that to define the $\ttb$-deformation in the Hamiltonian formalism one has to also specify how the constraints are deformed.

\section*{Acknowledgements}\noindent
We would like to thank Reiko Liu, Shi-Lin Zhu, Florian Loebbert and Sergey Frolov  for valuable discussions and comments.
The work is in part supported by NSFC Grants No. 11325522, No. 11735001, No. 11947301 and No. 12047502. J. T. is also supported by the UCAS program of special research associate and by the internal funds of the KITS.

\appendix

\section{Perturbative Lagrangian}\noindent
Expanding the $\ttb$-deformed Lagrangian by $\lambda$, $\mathcal{L}=\sum \lambda^n \mathcal{L}_n$, one can get a recursion relation from the flow equation \eqref{ttbareq},
\begin{widetext}
\be
\mathcal{L}_{i+1}=&\f{1}{i+1}\sum_{k=0}^{i}\[-\mathcal{L}_k\mathcal{L}_{i-k}+\mathcal{L}_k\vec{\beta}\cdot\f{\p \mathcal{L}_{i-k}}{\p \vec{\beta}}+\mathcal{L}_{i-k}\vec{\alpha}\cdot\f{\p \mathcal{L}_{k}}{\p\vec{\alpha}}-\(\f{\p \mathcal{L}_{k}}{\p \vec{x}}\cdot\vec{\alpha}\)\(\f{\p \mathcal{L}_{i-k}}{\p \vec{y}}\cdot\vec{\beta}\)+\(\f{\p \mathcal{L}_{i-k}}{\p \vec{\alpha}}\cdot\vec{\beta}\)\(\f{\p \mathcal{L}_{k}}{\p \vec{\beta}}\cdot\vec{\alpha}\)\].
\ee\end{widetext}
For one complex scalar, by the initial condition,
\be
\mathcal{L}_{0}=-V + X + Y
\ee
one can get the first several $\mathcal{L}_{k}$'s,
\be
\mathcal{L}_{1}=&-2 i A - (V - X - Y) (V + Y) \\
\mathcal{L}_{2}=&B - V^3 + V^2 X - V Y^2 - 2 i A (X + 2 Y)\\& + Y (X^2 + 3 X Y + 2 Y^2) \\
\mathcal{L}_{3}=&-4 A^2 - V^4 + V^3 X + B (X + 2 Y) - V Y^2 (3 X + 4 Y)\\& -
 2 i A (X^2 + 6 X Y - 2 (V - 3 Y) Y)\\& +
 Y (X^3 + 6 X^2 Y + 10 X Y^2 + 5 Y^3)\\
\mathcal{L}_{4}=&-V^5 + V^4 X + 2 V^2 Y^3 + B (X^2 + 6 X Y + 6 Y^2) \\&+
 Y (X^4 + 10 X^3 Y + 30 X^2 Y^2 + 35 X Y^3 + 14 Y^4) \\&+
 4 A^2 (V - 3 (X + 2 Y))\\& - V Y (2 B + Y (6 X^2 + 20 X Y + 15 Y^2))\\& -
 2 i A (2 B + (X + 2 Y) (X^2 + 10 X Y + 2 Y (-3 V + 5 Y))) \label{complex}
\ee

\section{Difference between two deformation formalisms}\noindent
\label{Diff}
To illustrate the difference between the deformations in the Lagrangian formalism and the Hamiltonian formalism, let us consider the well studied $\ttb$-deformed free scalar theory whose action was found in \cite{Cavaglia:2016oda} to take Nambu-Goto form
\be \label{L1}
\mathcal{L}_\lambda=-\frac{1}{\lambda}\left(\sqrt{1+\lambda (\p_x \phi)^2-\lambda (\p_t \phi)^2}-1\right).
\ee
Taking the derivative with respect to $\p_t \phi$ we can obtain the conjugate momentum
\be
\Pi=\frac{\p_t \phi}{\sqrt{1+\lambda (\p_x \phi)^2-\lambda(\p_t \phi)^2}}.\label{H2}
\ee
The standard Legendre transformation leads to the deformed Hamiltonian
\be \label{H1}
H_\lambda=\frac{1}{\lambda}\left( \sqrt{1+\lambda(\Pi_\lambda^2+(\p_x \phi)^2)+\lambda^2 \Pi_\lambda^2(\p_x \phi)^2}-1\right).
\ee
Now let us compare the first-order expansion of \eqref{L1} and \eqref{H1}:
\be\label{L2}
\mathcal{L}_\lambda=&\frac{1}{2}\left((\p_t \phi)^2-(\p_x \phi)^2\right)+\frac{1}{8}((\p_t \phi)^2-(\p_x \phi)^2)^2 \lambda+\mathcal{O}(\lambda^2)\\&\sim \mathcal{L}_0+\lambda \,\mbox{det}(T_{\mu\nu }),\\
\mathcal{H}_\lambda=&\frac{1}{2}\left(\Pi^2+(\p_x \phi)^2\right)-\frac{1}{8}((\Pi)^2-(\p_x \phi)^2)^2 \lambda+\mathcal{O}(\lambda^2).
\ee
The first-order correction of $H_\lambda$ can be written as
\be \label{H3}
\mathcal{H}_\lambda=\mathcal{H}_0-\lambda \,\mbox{det} (T_{\mu\nu })
\ee
if and only if we use the zeroth-order result of $\Pi$ in \eqref{H2}. It implies that if one wants to find the first-order expansion of the Hamiltonian from the Lagrangian via Legendre transformation, he has to rewrite the time derivative terms in terms of the conjugate momentum first, then perform the expansion while keep the conjugate momentum. However in the non-relativistic model we are studying, this can not be done. This fact leaves us with  two possible choices: one may take \eqref{H3} as the first order correction to the Hamiltonian or one may take the Legendre transformation of \eqref{L2} as the first order correction to the Hamiltonian. We have shown explicitly that these two choice give rise to different results.

\section{Brillouin-Wigner Perturbation}\noindent
In the textbook\cite{Shankar}, Rayleigh-Schr{\"o}dinger perturbation theory has been discussed. However, it is too complicated to go to a higher order. If we want to know the correction of the energy, we have to calculate the energy and wavefunctions simultaneously. And its physical meaning is not clear. In  contrast, the Brillouin-Wigner perturbation is easy to calculate to higher orders. One can calculate the energy alone without knowing the wavefunctions. It is the Feynman diagrams in quantum mechanics. One can find a good lecture in \cite{Sun:2016}.\par
For a theory whose Hamiltonian is $H=H_0+\lambda H'$, where $\lambda$ is small. The energy of the $k$-th eigenstate is given by
\be \label{BW}
E_{n}=&E_{n}^{(0)}+\lambda H'_{\mathrm{nn}}+\lambda^{2} \frac{H'_{n m} H'_{m n}}{E_{n}-E_{m}^{(0)}}\\&+\lambda^{3} \frac{H'_{n m} H'_{m m^{\prime}}H'_{m^{\prime} n}}{\left(E_{n}-E_{m}^{(0)}\right)\left(E_{n}-E_{m^{\prime}}^{(0)}\right)}+...
\ee
where
\be
H'_{\mathrm{ij}}=\left\langle\psi_{i}^{(0)}\left|H^{\prime}\right| \psi_{j}^{(0)}\right\rangle.
\ee
The superscript $(0)$ denotes the zeroth correction and $\psi_{i}$ denotes the $i$-th eigenstate. One need to sum over the repeated indices. Eq.
\eqref{BW} is an equation of energy $E_n$ alone. One can use the iteration to get the perturbative formula of the energy, which is just Rayleigh-Schr{\"o}dinger perturbation formula. By solving the equation, one can get the exact energy spectrum in principle. However, the equation has infinite number of terms and one cannot solve it in most cases. Fortunately, in the case at hand, we can solve it exactly.\par
There is a physical interpretation for Brillouin-Wigner perturbation. Eq. \eqref{BW} could be understood as the summation of Feynman diagrams: $H'_{ij}$ denotes the vertex and $\f{1}{E_n-E_m^{(0)}}$ denotes the propagator. For example,
\begin{widetext}
\be
  &H'_{n m}\frac{1}{E_{n}-E_{m}^{(0)}}H'_{m m^{\prime}}\f{1}{E_{n}-E_{m^{\prime}}^{(0)}}H'_{m^{\prime} n} =
  \vcenter{\includegraphics[scale=0.25]{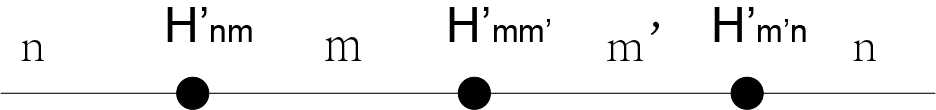}}   \\
\ee
\begin{equation}
  E_{n} =
  \vcenter{\includegraphics[scale=0.4]{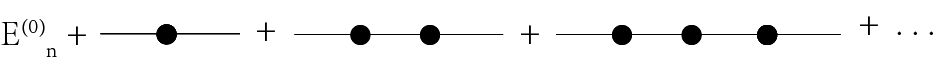}}  \\
\end{equation}
\end{widetext}

\section{Towards the second order deformation}\noindent
\label{second}
Expanding the exact deformed Lagrangian to the second-order of $\lambda$ gives
\be
\mathcal{L}=\mathcal{L}_0+\lambda \mathcal{L}_1+\lambda^2\mathcal{L}_2
\ee
with \eqref{FirstLag} and
\be
&\mathcal{L}_2=-2\p_x\phi^{\dagger2}\p_x\phi^2+i\p_x\phi \p_x\phi^\dagger\(\p_t\phi^\dagger\p_x\phi^2\phi^\dagger-\p_t\phi\p_x\phi^{\dagger2}\phi\)\\
&\qquad -\frac{1}{2}\p_t\phi\p_t\phi^\dagger\p_x\phi\p_x\phi^\dagger\phi\phi^\dagger+\f{i}{2}\p_x\phi^2\p_x\phi^{\dagger2}\(\p_t\phi\phi^\dagger-\p_t\phi^\dagger\phi\)\\
&\qquad+\frac{1}{4}\p_t\phi\p_t\phi^\dagger(\p_x\phi^{\dagger2}\phi^2+\p_x\phi^2\phi^{\dagger2}).
\ee
This Lagrangian density is not singular so we can solve $\p_t\phi$ and $\p_t\phi^\dagger$
\be
&\p_t\phi=\frac{4\Pi^\dagger+2 i(\phi+\lambda \phi^\dagger \p_x\phi^2+\lambda^2\phi \p_x\phi^2\p_x\phi^{\dagger2}-2\p_x\phi^3\phi^\dagger \p_x\phi^\dagger)}{\lambda^2(\p_x\phi \phi^\dagger -\p_x\phi^\dagger \phi)},\\
&\p_t\phi^\dagger=\frac{4\Pi-2 i(\phi^\dagger+\lambda \phi \p_x\phi^{\dagger2}+\lambda^2\phi^\dagger \p_x\phi^2\p_x\phi^{\dagger2}-2\p_x\phi^{\dagger3}\phi \p_x\phi)}{\lambda^2(\p_x\phi \phi^\dagger -\p_x\phi^\dagger \phi)}.
\ee
Therefore the Hamiltonian density is given by the Legendre transformation
\begin{widetext}
\be
\mathcal{H}=&\p_t \phi \Pi+\p_t \phi^\dagger \Pi^\dagger-\mathcal{L}\\
=&\frac{(2\Pi^\dagger+i \phi)(2\Pi-i \phi^\dagger)}{\lambda^2(\p_x \phi \phi^\dagger-\p_x\phi^\dagger\phi)^2}+\frac{i\p_x\phi^2\phi^\dagger(2\Pi-i\phi^\dagger)-i\p_x\phi^{\dagger2}\phi(2\Pi^\dagger+i\phi)}{\lambda(\p_x \phi \phi^\dagger-\p_x\phi^\dagger\phi)^2}\\
&+\frac{2i\Pi(\phi\p_x\phi \p_x\phi^\dagger-2\p_x\phi^2\phi^\dagger)-2i\Pi^\dagger(\phi^\dagger \p_x\phi\p_x\phi^\dagger-2\phi\p_x\phi^{\dagger2})-\p_x\phi^2\phi^{\dagger2}+\phi\phi^\dagger \p_x\phi\p_x\phi^\dagger-\p_x\phi^{\dagger2}\phi^2}{(\p_x \phi \phi^\dagger-\p_x\phi^\dagger\phi)^2}\\
&-\frac{2\lambda \phi\phi^\dagger\p_x\phi^3\p_x\phi^{\dagger3}}{(\p_x \phi \phi^\dagger-\p_x\phi^\dagger\phi)^2}+\frac{\lambda^2 \phi\phi^\dagger\p_x\phi^4\p_x\phi^{\dagger4} }{(\p_x \phi \phi^\dagger-\p_x\phi^\dagger\phi)^2}+\mathcal{O}(\lambda^3).
\ee
\end{widetext}
Because the Hamiltonian density is singular at $\lambda=0$ so the perturbation theory is not applicable. The appearance of a non-trivial denominator is also problematic. To proceed one may introduce proper constraints as before to reduce the Hamiltonian system even though it is ad hoc.  Based on our analysis of the first-order deformation, it seems that the most obvious constraints  are
\be
&\chi_1=\Pi-\frac{\p \mathcal{L}}{\p \p_t\phi}=0,\quad \chi_2=\Pi^\dagger-\frac{\p \mathcal{L}}{\p \p_t\phi^\dagger}=0.
\ee
Imposing these two constraints and using the equation of motion to get rid of $\p_t\phi$ and $\p_t\phi^\dagger$ we end up with a Hamiltonian system:
\begin{widetext}
\be \label{Ham2}
\mathcal{H}_2&=\p_x\phi\p_x\phi^\dagger-\lambda (\p_x \phi\p_x\phi^\dagger)^2+\frac{\lambda^2}{4}\( 8[\p_x \phi\p_x\phi^\dagger]^3+\p^2_x\phi\p^2_x\phi^\dagger [\p_x\phi\phi^\dagger-\p_x\phi^\dagger \phi]^2   \),\\
\chi_1&=\Pi-\frac{i \phi^\dagger}{2}-\frac{i\lambda\phi \p_x\phi^{\dagger2}}{2}+\frac{i\lambda^2}{4}\Big\{ (\p_x\phi \phi^\dagger-\p_x\phi^\dagger\phi)^2\p^2_x\phi^\dagger-2 (\p_x\phi\p_x\phi^\dagger)^2\phi^\dagger+4 \p_x\phi\phi(\p_x\phi^\dagger)^3\Big\} \\
&\equiv\chi_1^{(0)}+\lambda\chi_1^{(1)}+\lambda^2\chi_1^{(2)},\\
\chi_2&=\Pi^\dagger+\frac{i \phi}{2}+\frac{i\lambda\phi^\dagger \p_x\phi^{2}}{2}-\frac{i\lambda^2}{4}\Big\{ (\p_x\phi \phi^\dagger-\p_x\phi^\dagger\phi)^2\p^2_x\phi-2 (\p_x\phi\p_x\phi^\dagger)^2\phi+4\p_x\phi^\dagger\phi^\dagger(\p_x\phi)^3\Big\} \\
&\equiv\chi_2^{(0)}+\lambda\chi_2^{(1)}+\lambda^2\chi_2^{(2)}.\\
\ee
\end{widetext}
Using the Dirac-Bergmann Algorithm one can find
\be
&[\phi(x),\phi^\dagger(y)]_{DB}=\delta(x-y)\\&-\lambda C_1(x,y)-\lambda^2 \(C_2(x,y)-\int dx'  C_1(x,x')C_1(x',y)\),\\
&C_i(x,y)=[\Pi(y),\chi_2^{(i)}(x)]+[\chi_1^{(i)}(y),\Pi^\dagger(x)],\quad i=1,2.
\ee
Taking the ansatz of the canonical variables to be
\be
&\varphi=\phi-i\lambda \chi_2^{(1)}-i\lambda^2 \chi_2^{(2)}+\lambda^2 \phi_2,\\
&\varphi^\dagger=\phi^\dagger+i \lambda \chi_1^{(1)}+i\lambda^2 \chi_2^{(2)}+\lambda^2 \phi_2^\dagger,
\ee
and imposing the condition
\be
[\varphi(x),\varphi^\dagger(y)]_{DB}=\delta(x-y)
\ee
we end up with the equation for $\phi_2$ and $\phi_2^\dagger$
\be
&[\phi(x),\phi_2^\dagger(y)]+[\phi_2(x),\phi^\dagger(y)]-\frac{1}{4}\delta(x-y)(\p_x\phi(x))^2 (\p_x\phi^\dagger(x))^2\\&+\p_x\p_y\(\delta(x-y)\p_x\phi^\dagger \phi^\dagger \p_x\phi \phi\)=0.\\
\ee
Unfortunately we do not find local solutions to this equation.

\newpage

\clearpage

\end{document}